\documentclass{camera}

\usepackage{graphicx}  
\usepackage{amssymb}
\usepackage{amsmath}

\begin{document}

\title{Asymptotic solutions of pseudodifferential wave equations}

\author{Omar Maj}

\organization{INFM, Physics Department ``A.~Volta'', University of Pavia, Pavia, Italy}

\maketitle

\begin{abstract}
The aim of this paper is to give an account of some applications of pseudodifferential calculus for solving linear wave equations in the limit of high frequency/short wavelength waves. More specifically, on using as a benchmark the case of electromagnetic waves propagating in a cold isotropic slowly space- and time-varying plasma, it is shown that, in general, linear plasma waves are governed by pseudodifferential operators. Thereafter, the asymptotic techniques for solving the corresponding pseudodifferential wave equations are presented with emphasis on the \emph{paraxial} propagation of Gaussian wave trains in a cold isotropic plasma. Finally, it is addressed the unicity of the dispersion tensor in terms of  which the considered asymptotic solutions are determined.  
\end{abstract}

\section{Pseudodifferential wave equations: basic definitions and symbol calculus}
Let us consider a generic linear wave equation in the (abstract) form
\begin{equation}
\label{1}
\big(\hat{\bf D} \cdot \boldsymbol{\psi}\big)(x) = {\bf f}(x),
\end{equation}
where $\hat{\bf D}$ is a matrix of (linear) operators acting on a multi-component wavefield $\boldsymbol{\psi}(x)$ and ${\bf f}(x)$ an external drive (sources). Both $\boldsymbol{\psi}$ and ${\bf f}$ are defined on the $N$-dimensional real space, with $x=(x^{1},\ldots,x^{N})$ Cartesian coordinates, whereas $\hat{\bf D}$ is assumed to be a \emph{pseudodifferential} ($\Psi$DO) operator \cite{1,2}, that is, its action on the wavefield $\boldsymbol{\psi}$ amounts to the Fourier integral operator
\begin{equation}
\label{2}
\big(\hat{\bf D}\cdot \boldsymbol{\psi}\big)(x)= \frac{1}{(2\pi)^{N}}\int e^{ik\cdot (x-x')} {\bf d}(k,x,x')\cdot \boldsymbol{\psi}(x')d^{N}x'd^{N}k,
\end{equation}
where ${\bf d}(k,x,x')$ is a matrix of smooth functions, referred to as \emph{symbols}, that depend on the wavevector $k=(k_{1},\ldots,k_{N})$ as well as on the outcoming ($x$) and incoming ($x'$) variables. Moreover, symbols are characterized by the estimate \cite{1}
\begin{equation}
\label{3}
\partial_{k}^{\alpha}\partial_{z}^{\beta} {\bf d}(k,z) =O(\delta^{-m+|\alpha|}), \qquad\quad z\equiv (x,x')
\end{equation}
for $\delta = |kL|^{-1} \rightarrow 0$ (\emph{semiclassical limit}), where the scalelength $L$ characterizes the nonuniformity of the operator, being defined by $\partial_{z}^{\beta}{\bf d} \sim {\bf d}/L^{|\beta|}$; in equation (\ref{3}), $\alpha\equiv (\alpha_{1},\ldots,\alpha_{N})$ and $\beta\equiv (\beta_{1},\ldots,\beta_{2N})$ are multi-indices of dimension $N$ and $2N$, respectively, with, e.g., $\partial^{\alpha}_{k} \equiv \partial^{|\alpha|} / \partial k_{1}^{\alpha_{1}}\cdots\partial k_{N}^{\alpha_{N}}$ and $|\alpha| \equiv \sum_{i} \alpha_{i}$, whereas $m$ is a real number specifying the order of both the symbol and the corresponding operator. The condition (\ref{3}) is automatically satisfied if ${\bf d}$ is a polynomial in $k$ with smooth $x$-dependent (matrix-valued) coefficients, namely, ${\bf d}(k,x,x')\equiv{\bf d}(k,x)=\sum_{|\alpha|\leq m} {\bf d}_{\alpha}(x)k^{\alpha}$ with $m$ a positive integer, the corresponding $\Psi$DO amounting to the partial differential operator (PDO) $\hat{\bf D}=\sum_{|\alpha|\leq m} (-i)^{|\alpha|}{\bf d}_{\alpha}(x)\partial_{x}^{\alpha}$. It is worth noting that in this case the response $\big(\hat{\bf D}\cdot \boldsymbol{\psi}\big)(x)$ of the operator depends on the derivatives of the wavefield of order $\leq m$ only, i.e., $\hat{\bf D}$ is a \emph{local operator}; in order to account for \emph{nonlocal effects} in the response, polynomials (of \emph{integer} order $m$) have been generalized to symbols (of \emph{real} order $m$). 

As a consequence of the estimate (\ref{3}), symbols can be exploited in order to deal with the corresponding operators (\emph{symbol calculus}). With this aim it is convenient to perform, in equation (\ref{2}), the coordinate transformation $(x,x') \rightarrow(r\equiv qx+px', s\equiv x-x')$ with $q+p=1$, redefine the symbol according to $\tilde{\bf d}(k,r,s)\equiv {\bf d}(k, r+ps, r-qs)$ and expand with respect to $s$. 
As a result one gets the sequence of symbols ${\bf d}^{(q,p)}_{\ell}(k,x)\equiv\sum_{|\alpha|=\ell} \frac{i^{|\alpha|}}{\alpha!}\ \partial_{k}^{\alpha}\partial_{s}^{\alpha} \tilde{\bf d}(k,x,0)$ of decreasing order \big($=m-\ell$, cf., equation (\ref{3})\big) for $\ell=0,1,\ldots$, which can be \emph{resummed} \cite{1} to give the new symbol
\begin{equation}
\label{5}
{\bf d}^{(q,p)}(k,x) \sim \sum_{\ell=0}^{+\infty} {\bf d}^{(q,p)}_{\ell}(k,x),
\end{equation}
where the ``$\sim$'' means that, for every integer $n$, ${\bf d}^{(q,p)} - \sum_{\ell=0}^{n-1}{\bf d}_{\ell}^{(q,p)}$ is of order $m-n$. The new symbol (\ref{5}) is referred to as the \emph{reduced symbol} of the operator since it depends on a single $x$-variable rather than on both $x$ and $x'$. In the semiclassical limit $\delta =|kL|^{-1} \rightarrow 0$, the series (\ref{5}) is asymptotically convergent and for practical applications only the first two terms are significant. The advantage of the reduced symbol is that one can replace $\hat{\bf D}$ by any $\Psi$DO $\hat{\bf D}^{(q,p)}$ corresponding to ${\bf d}^{(q,p)}(k,qx+px')$, which depends on $(x,x')$ through the convex combination $qx+px'$ only. More specifically, the remainder $\hat{\bf R}\equiv \hat{\bf D}-\hat{\bf D}^{(q,p)}$ is a $\Psi$DO of order $m=-\infty$, i.e., the corresponding symbol ${\bf d} -{\bf d}^{(q,p)}$ satisfies the estimate (\ref{3}) for any order $m$, thus $\hat{\bf R}$ can be neglected in the semiclassical limit $\delta =|kL|^{-1}\rightarrow 0$, being formally of order $\delta^{+\infty}$. 

In particular, two specific forms of the reduced symbol play a key role in physical applications:
\begin{itemize}
\item the \emph{left}-reduced symbol ${\bf d}^{(L)} \equiv {\bf d}^{(1,0)}$,
\item the \emph{Weyl} symbol ${\bf d}^{(W)} \equiv {\bf d}^{(\frac{1}{2},\frac{1}{2})}$;
\end{itemize}
the former is the simplest generalization of the symbol of a partial differential operator (cf., section 2), whereas the latter properly accounts for the nonuniformity of the medium \cite{2} (cf., section 3). 

For any asymptotic form it is possible to represent the composition of operators through a suitable product-rule for symbols on noting that, given two scalar $\Psi$DO $\hat{A}$ and $\hat{B}$ with symbols $a$ and $b$, repectively, their composition $\hat{C} = \hat{A}\  \hat{B}$ corresponds to the symbol $c(k,x,x')=a^{(1,0)}(k,x)b^{(0,1)}(k,x')$ where the first (second) factor has been represented by means of the left (right, i.e., $q=0$, $p=1$) symbol \cite{1}; on applying the reduction procedure to $c(k,x,x')$, cf., equation (\ref{5}), and converting the form of the symbols $a^{(1,0)}$ and $b^{(0,1)}$ (again by means of (\ref{5}) with ${\bf d}$ replaced by $a^{(1,0)}(k,x)$ and $b^{(0,1)}(k,x')$ as appropriate), one gets, to lowest significant orders, \cite{1,2}
\begin{subequations}
\label{8}
\begin{eqnarray}
\label{8a}
c^{(L)} \sim a^{(L)} b^{(L)} - i \frac{\partial a^{(L)}}{\partial k_{i}}\frac{\partial b^{(L)}}{\partial x^{i}} + \cdots,\\
\label{8b}
c^{(W)} \sim a^{(W)} b^{(W)} + \frac{i}{2}\big\{a^{(W)}, b^{(W)}\big\} + \cdots,
\end{eqnarray}
\end{subequations}
for the left and the Weyl symbol, respectively, with $\{a,b\}\equiv \frac{\partial a}{\partial x^{i}}\frac{\partial b}{\partial k_{i}} -\frac{\partial b}{\partial x^{i}}\frac{\partial a}{\partial k_{i}}$ being the Poisson brackets in the $x$-$k$ phase space. The product-rules (\ref{8}) can be readily generalized to matrix-valued symbols (on replacing the product of functions by the product of matrices) and constitute a key tool in symbol calculus as they allow to perform calculations with integral operators by means of straightforward algebraic and differential operation on smooth functions.

\section{$\Psi$DO in plasma physics: an example}
In plasma physics pseudodifferential operators are found on dealing with linear wave propagation as a consequence of the dispersion combined with space- and time-variations of the background plasma.

As an example of how symbol calculus can be applied, let us consider a cold isotropic plasma for which the constitutive relationship for electromagnetic waves is given in implicit form by the generalized Ohm's law which relates the induced current density ${\bf J}$ to the wave electric field ${\bf E}$, namely,
\begin{equation}
\label{3.1}
\frac{\partial}{\partial t}\ {\bf J} = \frac{\omega_{pe}^{2}}{4\pi}\ {\bf E}
\end{equation}
$\omega_{pe}^{2}({\bf r},t)$ being the squared electron plasma frequency which depends on both the point-location ${\bf r}$ and time $t$ through the electron density profile. Equation (\ref{3.1}) can be cast in the form $\hat{Q}\ {\bf J} = {\bf E}$ with $\hat{Q}$ the first order partial differential operator with total symbol $q(\omega,{\bf r},t) = -i\frac{4\pi \omega}{\omega_{pe}^{2}} = -i\frac{4\pi (\omega + i\nu)}{\omega_{pe}^{2}} -\frac{4\pi\nu}{\omega_{pe}^{2}} \equiv q_{+1} +q_{0}$, where $\nu=\nu({\bf r},t)\not =0$ is a zero order symbol playing the role of a fictitious collisional frequency. In the high frequency limit, the properties of the operator $\hat{Q}$ are dominated by the symbol $q_{+1}$ which, however, is not unique being defined apart from the arbitrary function $\nu$ (for sake of completeness let us mention that, for symbols that can be written as a sum of homogeneous functions, the order of homogeneity allows to distinguish uniquely the leading term \cite{1}; here, however, such a property is ignored as it does not hold in the general case). 

Let us search for a $\Psi$DO $\hat{P}$ such that
\begin{equation}
\label{3.2}
\hat{P} \ \hat{Q} - \hat{I} = \hat{R}
\end{equation}
where $\hat{I}$ is the identity operator (with symbol $1$) and $\hat{R}$ a remainder of order $-\infty$. A solution $\hat{P}$ of equation (\ref{3.2}) is an approximate inverse (\emph{parametrix}) of $\hat{Q}$, thus, one gets ${\bf J} + \hat{R}\ {\bf J} = \hat{P}\ {\bf E}$ where the remainder $\hat{R}\ {\bf J}$ can be neglected in the semiclassical limit with the result that ${\bf J}({\bf r},t) \sim \big(\hat{P}\ {\bf E}\big)({\bf r},t)$.

The needed solution of equation (\ref{3.2}) is obtained on exploiting the product-rule (\ref{8a}) for left-reduced symbols which yields
\begin{equation}
\label{3.3}
p_{-1}q_{+1} + p_{-1}q_{0} + p_{-2}q_{+1} + i \frac{\partial p_{-1}}{\partial \omega}\ \frac{\partial q_{+1}}{\partial t} + \cdots =1,
\end{equation}
where $x=({\bf r},ct)$ and $k=({\bf k},-\omega/c)$, $p^{(L)} \equiv p_{-1} + p_{-2} + \cdots$ being the left symbol of the parametrix $\hat{P}$. Equation (\ref{3.3}) can be solved iteratively with the result that, to lowest significant orders,
\begin{subequations}
\label{3.4}
\begin{align}
\label{3.4a}
& p_{-1}(\omega,{\bf r},t) = \frac{1}{q_{1}(\omega,{\bf r},t)} = \frac{i\omega_{pe}^{2}}{4\pi (\omega +i\nu)},\\
\label{3.4b}
& p_{-2}(\omega,{\bf r},t)= \frac{\omega_{pe}^{2}}{4\pi(\omega +i\nu)^{2}} \left[\frac{2}{\omega_{pe}}\frac{\partial \omega_{pe}}{\partial t} -\nu\right]. 
\end{align}
\end{subequations}
In particular, one should note that $p_{-1}$ is the same as the conductivity of a stationary plasma with the time-dependence added adiabatically (\emph{adiabatic approximation}) whereas the next order term $p_{-2}$ explicitly accounts for the time variations of the plasma frequency. The induced current is related to the electric displacement by $\partial_{t}\big({\bf D} -{\bf E}\big) = 4\pi {\bf J}$ which can be solved for ${\bf D}$ on noting that the parametrix of $\partial_{t}$ corresponds to the symbol $i(\omega +i\nu)^{-1} - \nu(\omega +i\nu)^{-2} +\cdots$ and applying again the product-rule (\ref{8a}) with the result that ${\bf D}\sim \hat{\mathcal{E}}\ {\bf E}$, the dielectric operator $\hat{\mathcal{E}}$ being a $\Psi$DO of order zero with symbol \big(cf., equations (\ref{3.4})\big)
\begin{equation}
\label{3.5}
\varepsilon(\omega, {\bf r},t) = 1 - \frac{\omega_{pe}^{2}}{(\omega +i\nu)^{2}}\left[ 1- \frac{2i}{\omega + i\nu} \Big(\frac{1}{\omega_{pe}}\frac{\partial \omega_{pe}}{\partial t} - \nu \Big) + \cdots \right].
\end{equation}
It is worth noting that the symbol $\nu({\bf r},t)$ has no effects in the high frequency regime (as can be shown on Taylor-expanding with respect to $\nu/\omega\ll 1$), consistently with the considered generalized Ohm's law (\ref{3.1}) which does not accounts for collisions; indeed, $\nu$ has been introduced to deal the zero of $q$, though also a different (more formal) approach exists \cite{1} which is based on smoothly cutting off the symbol near $\omega=0$.

Finally, the resulting wave equation for the electric field reads
\begin{equation}
\label{3.6}
\frac{1}{c^{2}}\frac{\partial^{2}}{\partial t^{2}} \Big(\hat{\mathcal{E}}\ {\bf E}\Big) + \nabla\times(\nabla\times {\bf E})=0,
\end{equation}
which follows directly from Maxwell's equations (with no external drive).

The foregoing calculation of the response of a non-magnetized plasma exemplifies a general procedure that can be applied to more general cases (including, e.g., the effects of the magnetic field) and proves that the wave equation for electromagnetic waves in plasmas is, in general, a pseudodifferential equation.

\section{Semiclassical asymptotics of $\Psi$DO}
A general solution of the inhomogeneous wave equation (\ref{1}) amounts to the sum of a solution of the corresponding homogeneous problem, $\hat{\bf D}\cdot\boldsymbol{\psi}=0$, plus a particular solution of the whole inhomogeneous equation. 

{\bf The geometrical optics solution.} Let us first address the homogeneous problem and search for (asymptotic) solutions in the form (\emph{eikonal ansatz}) \cite{2,3}
\begin{equation}
\label{9}
 \boldsymbol{\psi}(x) = {\bf e}(x) A(x) e^{iS(x)},
\end{equation}
${\bf e}(x)$ and $A(x)$ being the slowly varying \big(i.e., $\partial_{x}({\bf e},A)\sim ({\bf e},A)/L$ with $L$ defined in (\ref{3})\big) polarization (unit) vector and complex amplitude, respectively, whereas the phase $S(x)$, referred to as the \emph{real} eikonal function, defines the local wavevector by $k(x)\equiv \partial_{x}S(x)$ which is ordered according to $|k(x)L| \sim \delta^{-1} (\rightarrow +\infty$ in the semiclassical limit). On substituting (\ref{9}) into (\ref{1}) with ${\bf f}=0$ and expanding with respect to $\delta$ one gets \cite{2}
\begin{multline}
\label{10}
{\bf d}(k,x,x)\cdot \big({\bf e}A\big) -i\frac{\partial {\bf d}(k,x,x)}{\partial k_{i}}\cdot \frac{\partial \big({\bf e}A\big)}{\partial x^{i}} -\frac{i}{2}\frac{\partial^{2}S(x)}{\partial x^{i}\partial x^{j}}\frac{\partial^{2}{\bf d}(k,x,x)}{\partial k_{i}\partial k_{j}}\cdot \big({\bf e}A\big) \\- i \frac{\partial^{2}{\bf d}(k,x,x)}{\partial k_{i}\partial x'^{i}}\cdot\big({\bf e}A\big) = O(\delta^{-m+2}),
\end{multline}
where $k=k(x)$, i.e., the eikonal ansatz (\ref{9}) allows to transform a pseudo-differential wave equation into a partial differential equation which has the form of an asymptotic series in $\delta$ and thus can be solved iteratively. More specifically, with reference to a weakly non-Hermitian media for which ${\bf d} = {\bf d}_{H} +i{\bf d}_{A}$ with ${\bf d}_{A}=O(\delta^{-m+1})$, the lowest order approximation of (\ref{10}) leads to ${\bf D}\big(\partial_{x}S(x),x\big)\cdot {\bf e}(x)=0$, ${\bf D}(k,x) \equiv {\bf d}_{H}(k,x,x)$ being the \emph{local} dispersion tensor. As far as mode conversion can be neglected \cite{3}, the polarization vector amounts to
\begin{subequations}
\label{11}
\begin{equation}
\label{11a}
{\bf e}(x) = {\bf e}\big(\partial_{x}S(x),x\big)
\end{equation}
with ${\bf e}(k,x)$ being an eigenvector of the dispersion tensor, corresponding to the eigenvalue $D(k,x)$. Thereafter, the phase $S(x)$ is determined by the first order partial differential equation (referred to as the \emph{eikonal equation}) 
\begin{equation}
\label{11b}
D\big(\partial_{x}S(x),x\big)=0,  
\end{equation}
which is analogous to the Hamilton-Jacobi equation for a Hamiltonian (mechanical) system in the $x$-$k$ phase space with $D(k,x)$ as Hamiltonian function \cite{2,3}, so that the phase  $S(x)$ can be computed as the action relevant to such a Hamiltonian system. To the next order, equation (\ref{10}) yields the transport equation for the (squared) amplitude, namely,
\begin{equation}
\label{11c}
\frac{\partial}{\partial x^{i}}\left[\frac{\partial D}{\partial k_{i}} |A|^{2}\right] = 2\gamma_{A} |A|^{2},\quad \gamma_{A} \equiv {\bf e}{\bf e}^{*}:\left[{\bf d}_{A} +\frac{1}{2}\Big( \frac{\partial^{2}{\bf d}_{H}}{\partial k_{i}\partial x^{i}} - \frac{\partial^{2}{\bf d}_{H}}{\partial k_{i}\partial x^{\prime i}}\Big) \right]
\end{equation}
\end{subequations}
where $k=k(x)$ and $x'=x$. Equation (\ref{11c}) is a continuity equation for the flux vector $\frac{\partial D}{\partial k} |A|^{2}$ which points toward the velocity $V^{i} =\frac{\partial D}{\partial k_{i}}$ tangent to the trajectories of the equivalent mechanical system in the configuration ($x$) space, referred to as the geometrical optics (GO) rays, $\gamma_{A}$ being the absorption coefficient. As a consequence, equation (\ref{11c}) can be reduced to an ordinary differential equations along the GO rays and solved in parallel to the Hamilton's equations with limited computational efforts.  

It should be noted that the form (\ref{11}) of the GO equations holds independently on whether one makes use of the exact symbol ${\bf d}(k,x,x')$ or any of its reduced forms ${\bf d}^{(q,p)}(k,qx+px')$; in the latter case the dispersion tensor amounts to ${\bf D}(k,x)={\bf d}_{H}^{(q,p)}(k,x)$, whereas the absorption coeffcient reads $\gamma^{(q,p)}_{A}={\bf e}{\bf e}^{*}:\left[{\bf d}^{(q,p)}_{A}-\big(p-\frac{1}{2}\big)\frac{\partial^{2}{\bf D}}{\partial k_{i}\partial x^{i}}\right]$ which, in particular, takes the simplest form if the Weyl symbol is used, namely, $p=\frac{1}{2}$ and  $\gamma_{A}^{(\frac{1}{2},\frac{1}{2})}={\bf e}^{*}\cdot {\bf d}_{A}^{(\frac{1}{2},\frac{1}{2})} \cdot {\bf e}$. However, the polarization vector, the Hamiltonian function and the absorption coefficient are different depending on the choice of the asymptotic form of the wave operator (\ref{2}). One can prove that different formulations yield GO solutions which differ for $O(\delta)$-corrections \cite{4}.

{\bf The Wigner-Weyl kinetic formalism}. The GO solution discussed above exhibits two limitations: it has a local validity and does not apply to the inhomogeneous wave equation. To some extent, such limitations can be overcome on representing the wavefield in the $x$-$k$ phase space. With this aim, the wave equation (\ref{1}) is written in the quadratic form \cite{2}
\begin{equation}
\label{4.1}
\hat{\bf D} \cdot (\boldsymbol{\psi}\boldsymbol{\psi}^{\dagger}) = ({\bf f}{\bf f}^{\dagger})\cdot (\hat{\bf D}^{\dagger})^{-1},
\end{equation}
where $(\hat{\bf D}^{\dagger})^{-1}$ can be obtained by means of symbol calculus whereas $(\boldsymbol{\psi}\boldsymbol{\psi}^{\dagger})$ and $({\bf f}{\bf f}^{\dagger})$ are projection operators. On assuming that both $(\boldsymbol{\psi}\boldsymbol{\psi}^{\dagger})$ and $({\bf f}{\bf f}^{\dagger})$ are $\Psi$DO, one can apply the product-rule (\ref{8b}) for the corresponding Weyl symbols with the result that equation (\ref{4.1}) is transformed into \cite{2}
\begin{equation}
\label{4.2}
{\bf d}^{(W)} \cdot {\bf S} + \frac{i}{2}\big\{ {\bf d}^{(W)}, {\bf S}\big\} +\cdots = {\bf N} \cdot ({\bf d}^{(W)\dagger})^{-1} + \cdots,
\end{equation}
${\bf S}$ and ${\bf N}$ being the Weyl symbols of $(\boldsymbol{\psi}\boldsymbol{\psi}^{\dagger})$ and $({\bf f}{\bf f}^{\dagger})$, respectively. Equation (\ref{4.2}) can be solved iteratively as for the corresponding equation (\ref{10}) in the GO method. In particular, for weakly non-Hermitian $\big({\bf d}^{(W)}_{A}=O(\delta^{-m+1})\big)$ and weakly driven \big(${\bf N} \cdot ({\bf d}^{(W)\dagger})^{-1} = O(\delta^{-m+1})\big)$ media and far from mode conversion regions, one has ${\bf S}(k,x) = W(k,x) {\bf e}(k,x){\bf e}^{*}(k,x)$  with \cite{2}
\begin{subequations}
\label{4.3}
\begin{align}
\label{4.3a}
& D(k,x)W(k,x)=0,\\
\label{4.3b}
& \big\{W, D\big\} = 2\gamma_{A}^{(W)} W + \frac{2i({\bf e}^{*}\cdot {\bf N}\cdot {\bf e})}{D-i\gamma_{A}^{(W)}},
\end{align}
\end{subequations}
$D(k,x)$ and $\gamma_{A}^{(W)} \equiv \gamma_{A}^{(\frac{1}{2},\frac{1}{2})}$ being, respectively, the eigenvalue of ${\bf d}_{H}^{(W)}$ corresponding to the eigenvector ${\bf e}$ and the absorption coefficient in the Weyl form \big(cf., comments after equation (\ref{11c})\big). Equation (\ref{4.3b}) has the form of a kinetic equation in the $x$-$k$ phase space for the scalar function $W(k,x)$, referred to as the \emph{Wigner function}, with the constraint (\ref{4.3a}) representing the local dispersion relation $D(k,x)=0$. The Wigner function is related to the wavefield intensity through the identity $|\boldsymbol{\psi}(x)|^{2}= \int\frac{d^{N}k}{(2\pi)^{N}} W(k,x)$ \cite{2}, however, one can show that the approximations underlying the asymptotic series (\ref{4.2}) are such that the solution of the equations (\ref{4.3}) allows to compute the wavefield intensity averaged over the large scalelength $\ell (\gg 2\pi/|k|)$, $\langle|\boldsymbol{\psi}(x)|^{2}\rangle_{\ell}$, rather than the exact intensity $|\boldsymbol{\psi}(x)|^{2}$ \cite{5}.

Let us note that the Poisson brackets in (\ref{4.3b}), which are due to the specific product-rule (\ref{8b}) for Weyl symbols, endows the $x$-$k$ space with the same Hamiltonian structure as in the GO technique, provided that the Weyl formulation is used.

{\bf Paraxial solutions of pseudodifferential wave equations.} In the GO method, both the polarization vector ${\bf e}$ and the complex amplitude $A$ have been assumed to be slowly varying on the scalelength $L$ of the medium nonuniformity. Such an ansatz is justified for the polarization ${\bf e}$ far from mode conversion regions \big(cf., equation (\ref{11a})\big), but \emph{not} for the amplitude $A$ that, being transported along the GO rays, can develope its own scalelength $w$, called \emph{beamwidth} in analogy to the case of wavebeams, namely, $\partial_{x}A \sim A/w$. Depending on the geometry of the bundle of GO rays, the beamwidth $w$ can be shorter that $L$, e.g., near focal points, and should be explicitly accounted for in the asymptotic analysis. With this aim, quasi-optics methods \cite{6} have been developed, for the case of monochromatic electromagnetic wavebeams propagating in stationary, spatially nondispersive media for which the relevant wave equation exhibits local operators only. Among such quasi-optics techniques, the complex geometrical optics (CGO) method \cite{6} is proved to be a benchmark on the basis of which other quasi-optics solutions can be derived \cite{7}; in particular, within the \emph{paraxial} approximation $(2\pi/|k|) \ll w \big(\sim \sqrt{2\pi L/|k|}\big) \ll L$, the CGO method reduces to the beam tracing (BT) technique \cite{8} which has the advantage of being particularly suited for numerical analysis. 

On combining the CGO method with the paraxial approximation, one can prove that the beam tracing method can be used to solve generic pseudodifferential wave equations provided that the range of nonlocal effects is finite \cite{5}.
Let us describe the general features of such BT solutions on the basis of the simple analytically-tractable case addressed in section 2; in particular, let us assume that the plasma is homogeneous and stationary with $\varepsilon(\omega,{\bf r},t)=n^{2}(\omega)$, $n(\omega)$ being the refractive index, and search for solution of (\ref{3.6}) in the form ${\bf E} = \hat{\bf y}\ \psi(x,t)$, the relevant equation for the scalar field $\psi(x,t)$ amounting to
\begin{equation}
\label{12}
\frac{1}{c^{2}}\frac{\partial^{2}}{\partial t^{2}}\big(\hat{N}^{2}\psi\big) -\frac{\partial^{2} \psi}{\partial x^{2}} =0,
\end{equation}
with $\hat{N}$ the $\Psi$DO corresponding to the refractive index $n(\omega)$.
The wavefield at the initial time $t=0$ is assumed to be a \emph{wave train} characterized by the wavelength $2\pi/k_{0}$ and by a Gaussian envelope with the initial $e^{-1}$-width $w_{0}$, namely, $\psi(x,t=0)=\psi_{0}(x) = A_{0} e^{-x^{2}/w_{0}^{2}}\cos(k_{0}x)$. It is worth noting that the exact solution of such a problem could be formally achieved by means of the Fourier transform, but the explicit calculation of the wavefield requires an integration in both the frequency $\omega$ and wavevector $k$ which, in general, is not straightforward both analytically and numerically. On the other hand, the BT method relys on a set of ordinary differential equations [5-8] the coefficients of which are readily obtained as derivatives of the symbol $D(k,\omega)=-\frac{\omega^{2}}{c^{2}}n^{2}(\omega) + k^{2}$ corresponding to the $\Psi$DO in (\ref{12}). The solutions of such equations yield the wavefield directly in the $(x,t)$ space-time, namely, \cite{5}
\begin{multline}
\label{13}
\psi(x,t) = A_{0}\Big[\frac{w_{0}}{w(t)}\Big]^{\frac{1}{2}} e^{-\frac{(x-v_{g}t)^{2}}{w^{2}(t)}} \\
\times\cos\Big[S_{0}(t) + k_{0}(x-v_{g}t) +\frac{k_{0}}{2R(t)}(x-v_{g}t)^{2} -\varphi(t) \Big],    
\end{multline}
which is shown in figure 1. Let us note that the BT solution (\ref{13}) is localized around the space-time curve $(x=v_{g}t,t)$ with $v_{g} = v_{p}/\big(1+\frac{1}{\omega}\frac{dn}{d\omega}\big)$ being the group velocity ($v_{p}= c/n$, the phase velocity), whereas the $e^{-1}$-width $w(t) = \big[1+\frac{4((1-r)v_{g}t)^{2}}{k_{0}^{2}w_{0}^{4}}\big]^{\frac{1}{2}} w_{0}$ increases throughout the propagation so that the Gaussian envelope spreads out depending on the adimensional parameter $r=\frac{v_{g}^{2}}{v_{p}^{2}}\big[1+\frac{4\omega}{n}\frac{dn}{d\omega} + \frac{\omega^{2}}{n^{2}}\big(\frac{dn}{d\omega}\big)^{2} + \frac{\omega^{2}}{n^{2}}\frac{d^{2}n}{d\omega^{2}}\big]$ which controls the dispersion effects (for a non dispersive medium $r=1$ and the BT solution (\ref{13}) reduces to the well-known solution of the d'Alembert wave equation). As a consequence of the spreading the wavefield, the amplitude $A_{0}\big[\frac{w_{0}}{w(t)}\Big]^{\frac{1}{2}}$ drops due to the conservation of the squared phase-averaged amplitude integrated with respect to $x$. The short-scale oscillations of the wavefield are accounted for by the $\cos$-term in (\ref{13}), the argument of which exhibits a rather fine structure: $S_{0}(t) = -i\omega t'$, with $t'=t-v_{g}t/v_{p}$ the retarded time, describes the temporal oscillations, the term $k_{0}(x-v_{g}t)$ is just the initial phase $k_{0}x$ shifted along the space-time curve $(x=v_{g}t,t)$ whereas the term $\frac{k_{0}}{2R(t)}(x-v_{g}t)^{2}$ accounts for the curvature of the phase-fronts (viewed as curves in the two-dimensional space-time) with $R(t)=(1-r)v_{g}t + \frac{k_{0}^{2}w_{0}^{4}}{4(1-r)v_{g}t}$ the radius of curvature; finally, the phase-shift $\varphi(t) = \arctan\big[\frac{2(1-r)}{k_{0}w_{0}^{2}} v_{g}t\big]$ is due to the effect of the envelope spreading (the analogous of the Gouy shift for wavebeams \cite{5, 7}). In equation (\ref{13}) any function of the frequency should be evaluated at $\omega = \omega_{0}$, with $\omega_{0}$ solution of the dispersion equation $D(k_{0},\omega_{0})=0$. 

\begin{figure}
\begin{center}
\begin{tabular}{cccc}
\put(200,20){\tiny $\omega_{0}t$}
\put(290,25){\tiny $x/w_{0}$}
\put(73,-10){\tiny (a)}
\put(245,-10){\tiny (b)} 
\includegraphics[bb =51 210 433 718, scale = .35]{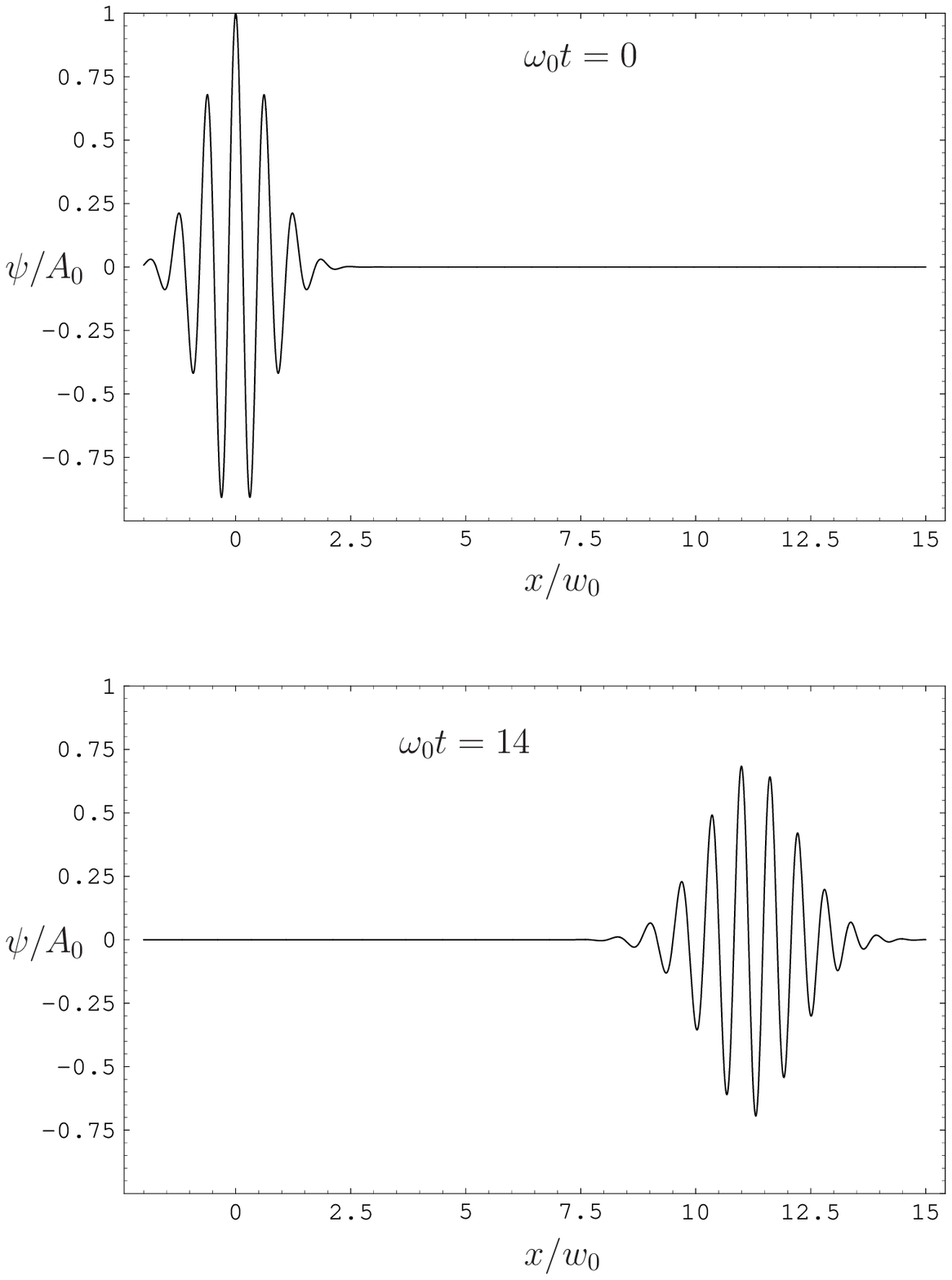}
& & &
\includegraphics[bb =100 0 332 175, scale = .65]{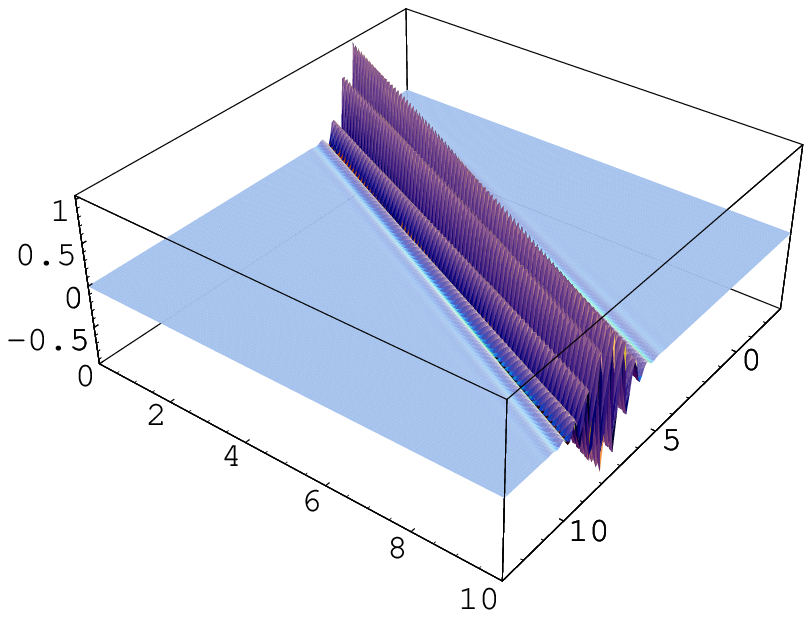}
\end{tabular}
\end{center}
\caption{Time evolution of a Gaussian wave train according to the BT solution (\ref{13}) ($r=0.2$ and $v_{g}/v_{p}=0.8$): (a) two constant-time sections showing the propagation and spreading of the wavefield and (b) the plot of the solution in the two-dimensional space-time showing the localization of the wavefield around the curve $(x=v_{g}t,t)$.}
\end{figure}

Let us note that several features of the BT solution (\ref{13}) are general. In particular, the wavefield can be written in form of a \emph{complex}-eikonal wavelet [6-8] $\psi(x,t)=\Re e\big\{ A(x,t) e^{i\bar{S}(x,t)} \big\}$ where both the complex amplitude $A(x,t)$ and the \emph{complex}-eikonal function $\bar{S}$ have been expanded in Taylor series with respect to $(x-v_{g}t)$ in a neighbourhood of the space-time curve $(x=v_{g}t,t)$, with time $t$ labelling the position along the curve and $(x-v_{g}t)$ being its distance from the considered point $(x,t)$ (\emph{paraxial expansion}), namely,
\begin{align*}
& A(x,t) = A_{0}(t) + O(\delta^{\frac{1}{2}}),\qquad A_{0}(t) = A_{0}\Big[\frac{w_{0}}{w(t)}\Big]^{\frac{1}{2}} e^{-i\varphi(t)},\\
& \bar{S}(x,t) = S_{0}(t) + k_{0}(x-v_{g}t) +\frac{1}{2}\left[\frac{k_{0}}{R(t)} + \frac{2i}{w^{2}(t)} \right] (x-v_{g}t)^{2} + O(\delta^{\frac{1}{2}}),
\end{align*}
which holds within the paraxial approximation with $|x-v_{g}t| \lesssim w(t) \sim \delta^{\frac{1}{2}}L$.
As a consequence, the whole wavefield around the reference curve (which in the general case amounts to a GO ray referred to as the \emph{reference ray}) is completely determined by a set of parameters, e.g., the amplitude $A_{0}(t)$, the phase $S_{0}(t)$, the width $w(t)$ and the curvature radius $R(t)$, that satisfy ordinary differential equations along the reference ray, so that one can find paraxial solutions of rather complicated wave equations (including nonlocal effects) by simple computational means.

\section{The intrinsic Hamiltonian structure}

It has been stressed that in the paraxial solution (\ref{13}) the wavefield is localized around a geometrical optics (GO) reference ray, which represents the trajectory of the whole wave train in the space-time. On the other hand, the reference ray depends on the adopted GO formulation \big(cf., comments after equation (\ref{11c})\big), so that different forms of the dispersion tensor ${\bf D}(k,x)$ corresponds to different reference rays which can deflect one from the other for long-enough propagation. Hence, one should wonder what GO formulation yields the reference ray around which the wavefield is actually localized \big(let us note that the symbol of the operator (\ref{12}) is uniquely defined, as follows from equation (\ref{5}), thus, in this case, there is no ambiguity in the definition of the reference ray\big). 

With this aim, it is worth noting that the symbol ${\bf d}$ of the wave operator (\ref{2}) is usually obtained by means of symbol calculus (e.g., as in section 2) and thus it amounts to an asymptotic series, namely, 
\begin{equation}
\label{30}
{\bf d}(k,x,x') \sim {\bf d}_{0}(k,x,x') + {\bf d}_{1}(k,x,x') + \cdots
\end{equation}
with ${\bf d}_{\ell}$ being of order $m-\ell$. Moreover, the leading term in the reduced symbol (\ref{5}), i.e., ${\bf d}_{0}^{(q,p)}(k,x)= {\bf d}_{0}(k,x,x)$, is an Hermitian tensor for weakly non-Hermitian media and \emph{does not depend on the adopted $(q,p)$-formulation}; for this reason, ${\bf d}_{0}$ is referred to as the \emph{principal symbol} of the operator. Therefore, in spite of separating the Hermitian and anti-Hermitian parts of the total symbol ${\bf d}$, one should make use of the asymptotic expansion (\ref{30}) into both equations (\ref{10}) and (\ref{4.2}) with the result that the dispersion tensor is determined by the \emph{unique principal symbol}, namely, ${\bf D}(k,x) \equiv {\bf d}_{0}(k,x,x)$, and yields the same polarization vector and Hamiltonian function in any GO formulation as well as in the Wigner-Weyl formalism. As for the absorption coefficient, one gets \big(cf., equation (\ref{11c})\big) 
\begin{multline*}
\gamma_{A} = {\bf e}{\bf e}^{*} : \left[ {\bf d}_{1,\ A} + \frac{1}{2}\Big(\frac{\partial {\bf d}_{0}}{\partial k_{i}\partial x^{i}} -\frac{\partial^{2}{\bf d}_{0}}{\partial k_{i}\partial x'^{i}}\Big)\right] \\
\equiv {\bf e}{\bf e}^{*}:\left[{\bf d}_{1,\ A}^{(q,p)} - \Big(p-\frac{1}{2} \Big) \frac{\partial^{2}{\bf D}}{\partial k_{i}\partial x^{i}} \right]= \gamma_{A}^{(q,p)},
\end{multline*}
for any $(q,p)$, where the asymptotic series (\ref{5}) has been accounted for toghether with (\ref{30}).

In conclusion, on separating the principal symbol it is possible to identify an \emph{intrinsic} dispersion tensor which does not depend on the considered form of the operator and thus yields the same polarization vector, Hamiltonian function and absorption coefficient in every formulations of the geometrical optics method as well as in the Wigner-Weyl kinetic formalism \cite{5}. However, such a separation is, in general, defined apart from the transformation ${\bf D} \rightarrow {\bf D} +{\bf g}$ with ${\bf g}$ a lower-order Hermitian symbol, e.g., the principal symbol $q_{+1}$ of the operator $\hat{Q}$ given in section 2 is defined on adding and arbitrary lower-order symbol $\nu$. Such an arbitrarity should be dealt with in term of a Gauge transformation \cite{3}.


{\small
\begin{thebibliography}{8}
\bibitem{1} JOSHI M., math-AP/9906155 Preprint 1999, and references therein.
\bibitem{2} McDONALD S. W., {\it Phys. Rep.} {\bf 158} (1988) 337.  
\bibitem{3} LITTLEJOHN R. G. and FLYNN W. G., {\it Phys. Rev. A} {\bf 44} (1991) 5239.   
\bibitem{4} BORNATICI M. and MAJ O., {\it Plasma Phys. Control. Fusion} {\bf 45} (2003) 1511.
\bibitem{5} MAJ O., PhD thesis, University of Milano Italy 2003.  
\bibitem{6} PEETERS A. G., PEREVERZEV G. V. and WESTERHOF E., in {\it Proc. of the 10th Joint Workshop on ECE and ECH} 1997, p.17 and references therein.
\bibitem{7} BORNATICI M. and MAJ O., {\it Plasma Phys. Control. Fusion} {\bf 45} (2003) 707.
\bibitem{8} PEREVERZEV G. V., in this proceedings 2004. 
\end{thebibliography}
}
\end{document}